\documentstyle[epsfig,,mathptm,subfigure,amsmath,referee]{mn}
\title[The SZ effect in Cl~0016+16] {Mapping of the
      SZ effect in the cluster Cl~0016+16 with the Ryle Telescope} 

\author[Keith Grainge et al.]
      {Keith Grainge, William F. Grainger, Michael E. Jones, R\"udiger
      Kneissl,\cr Guy G. Pooley, Richard Saunders.
      \\ \\ Astrophysics Group,
      Cavendish Laboratory, Madingley Road, Cambridge, CB3 0HE}

\date{Accepted  ;
      Received  ;
      in original form  }

\begin{document}

\maketitle

 \begin{abstract} We have mapped the high-redshift (z = 0.546) cluster
Cl~0016+16 
with the Ryle Telescope at 15 GHz. The Sunyaev--Zel'dovich decrement is
clearly detected, and resolved. We combine our data with an X-ray
image from {\sc Rosat}, and a gas temperature from {\sc Asca}
to estimate the Hubble Constant
$H_0=69^{+21}_{-16}~\rm km \ s^{-1} \ Mpc^{-1}$ 
 for an $\Omega_M=1.0$
cosmology or $H_0=84^{+25}_{-19}~\rm km \ s^{-1} \ Mpc^{-1}$
for $\Omega_M=0.3$ and $\Omega_\Lambda=0.7$.
\end{abstract} 

\begin{keywords} cosmic
microwave background -- galaxies: clusters: individual: Cl~0016+16 -- galaxies:
intergalactic medium -- radiative transfer -- X-rays: sources \end{keywords}

\section{Introduction}

The cluster Cl~0016+16 lies at a redshift of 0.546~\cite{DG}, and in X-rays is
one of the most luminous clusters known~\cite{Henry}. Despite its high redshift
it has a predominance of red galaxies (i.e. it defies the Butcher--Oemler
effect~\cite{BO}). There is also evidence that it is undergoing a merger; the
galaxy distribution is distinctly bimodal, and a map of the total mass derived
from weak lensing of background galaxies~\cite{Smail} shows several peaks,
although the X-ray emission appears somewhat smoother~\cite{BH}.

Cl~0016+16 was also one of the first clusters in which the Sunyaev--Zel'dovich
(SZ, Sunyaev \& Zel'dovich~\shortcite{sandz}) effect was
detected~\cite{Birkinshaw}. 
It has been subsequently mapped with the OVRO/BIMA array at
28--30~GHz~\cite{CJG96} and detected at 2.1~mm and 1.2~mm with the
Diabolo experiment at the IRAM-30~m telescope~\cite{D98}.
Here we present Ryle Telescope~(RT)
observations of the SZ effect in Cl~0016+16 and provide an
image of its two-dimensional structure. 
We then combine this data with the
{\sc Rosat} PSPC image and {\sc Asca} temperature data to calculate
the Hubble constant.

\section{SZ Observations with the Ryle Telescope}

The RT is an east--west aperture synthesis telescope
specifically designed to map low-surface-brightness structures in the
microwave background~\cite{Jones91}. Five of its eight antennas are used in
close-packed configurations to provide the high temperature sensitivity needed
to map the SZ effect. The system temperature at 15~GHz is 60~K, giving a flux
sensitivity in 12~h of 200~$\mu {\rm Jy \; beam^{-1}}$.  The RT correlator
provides 35 spectral channels of approximately 10~MHz bandwidth each
(7 synthesised 
channels in each of 5 IF sub-bands).

We have made many detections of the SZ effect with the
RT~\cite{J93,G93,JonesCapri,SaundersCapri}; we will first describe in
some detail the general observing procedure, and then discuss the Cl~0016+16
observations.

\subsection{General observing procedure}

Clusters are observed in runs of 12~hours with the telescope in one of the
compact configurations shown in table~\ref{conf}.
In the case of fields of declination less than $48^\circ$ 
with the RT in configuration~Ca or Cb,
the observation is curtailed when none of the baselines of the compact array
remain unshadowed.  
Due to the east-west nature of the RT, this results in a gap along the
{\it v}-axis in the aperture plane coverage. To fill this region we
also observe in either Cc array (for $24^\circ<\rm{dec}<48^\circ$) or Cd array
(for $16^\circ<\rm{dec}<24^\circ$). 
The pointing centre chosen for each observation is the
position of the peak of the X-ray emission from the cluster.  For each
observation, a bright ($>0.2$ Jy) nearby point source is chosen as a phase
calibrator and is observed for either five or ten 32-second samples
(depending on 
its brightness), for every 30 samples on-source.  A primary calibrator, either
3C48 or 3C286, is observed either before or after a run.

\subsection{Editing the data}

The {\sc Postmortem} package~~\cite{T91} is used for editing and calibrating of
RT data. The performance of the RF system is monitored by the use of a
modulated noise source at each feed-horn. 
Corrections to the data and their weights are applied to
take account of variations in system noise, mainly a result of weather
conditions. Exceptionally bad data are discarded, along with that
during winds in excess of 20~knots (marginal pointing) and during
slewing and settling while changing pointing centres, and for any
baseline actually, or close to being, shadowed.

\subsection{Primary calibration}

The absolute flux scale of the RT is fixed on 3C48 and 3C286. 
Flux densities of
1.70 and 3.50~Jy (Stoke's $I-Q$) are assumed for these
respectively~\cite{B77}.
We calculate a set of complex gains to apply to every sub-band and
channel of
each antenna from the primary calibrator observations. A least-squares
algorithm is used to derive the gains for each antenna from the data for each
baseline. This means that data from many baselines are used to calibrate each
antenna, and increases the signal-to-noise of the calibration. 
The amplitude scale is observed not to
drift significantly over the timescale of individual observations, and so is
calibrated with a single correction for each observation.  This primary
calibration is then applied to the data from the main observation.
Monitoring by the VLA has detected low level variability in both of
these sources on long time-scales~\cite{vlacalib}, and we apply a
separate correction for these changes.

\subsection{Phase calibration}

Antenna-based phase corrections are calculated for each of the 5- or
10-sample secondary calibrator observations, and interpolated to correct the
source phases. This removes phase drifts due to the atmosphere and local
oscillator distribution, which can be up to $360^{\circ}$ during a 12-hour run
for the longer baselines of the compact array. Since phase differences between
channels in the same sub-band are relatively constant, and are calibrated out
using the flux calibrator observations, we merge the visibilities of the
7~channels for each sub-band and smooth over all 5 or 10 samples within the
particular calibrator pointing in order to increase the signal-to-noise ratio
of the calibration.  We do not merge the secondary calibrator visibilities
over the sub-bands since we have found that during a 12-hour run the different
sub-bands can drift by up to ~$5^{\circ}$. Closure phase errors are calculated
when the antenna solutions are found, and are almost always consistent with
the thermal noise.  The data are flagged where this is not the case.
In order to check the secondary calibration the
visibilities are smoothed, their mean amplitudes and phases are examined and
the data for one representative channel are plotted out.  We pick out and flag
anomalous mean amplitudes, phases and their rms, and look for amplitude or
phase jumps caused by interference or possible hardware faults.

\subsection{Smoothing, weighting and interference clips}

The data are smoothed to an integration time appropriate to the resolution of
the array: 320~s for the most compact arrays Ca, Cb and Cc, and 96~s for array
Cd.  These smoothing periods were chosen to ensure that $R_a$, the fractional
decrease in the peak response to a point source is close to unity. $R_a$ is
given by (see e.g. Thompson~\shortcite{TMS86})
\begin{equation}
R_a \approx 1 - {1\over3} \left({{0.833 \omega_e \tau_a
r}\over{\theta_b}} \right)^2 , 
\end{equation}
where $\omega_e$ is the rotation rate of the Earth, $\tau_a$ is the averaging
time, $\theta_b$ is the FWHM of the synthesised beam, and $r$ is the distance
of the point source from the observation centre.  For the smoothing periods
used, a point source 6~arcminutes from the pointing centre (the position of
the first null of the RT primary beam response) will have $R_a \ge 0.99$ on
the longest baseline available in the respective configurations.  Any
low-level interference is removed by a simple clip at 0.25~Jy before
smoothing, and at 0.079~Jy after smoothing over 10~samples, or at 0.144~Jy
after smoothing over 3~samples. This corresponds to approximately 3~standard
deviations. The visibilities are weighted by their inverse variances,
calculated from the product of the system temperatures of the two antennas and
are then written out in {\sc fits} format to be transferred to {\sc
Aips}. As a final check of our calibration procedure,
 each day's data are mapped individually using the entire
{\it uv}-range 
and natural weighting, in order to give maximum signal to noise. We examine
these maps for artifacts due to poor calibration and interference which have
not been removed. 

\subsection{Data analysis}

Before a map of the SZ effect can be made, the effects of discrete sources on
the data have to be removed. The clusters we have studied have been
pre-selected to be radio-quiet, but may contain several radio sources with
flux densities of up to a few mJy. Rather than attempt to identify and
determine the fluxes of these sources in the map plane through
{\sc Clean}ing the long baseline data as has been our practice
previously~\cite{G93,J93}, we now fit directly to the visibilities
in the aperture plane using the program {\sc Fluxfitter}.
 This procedure is explained in greater depth in
Grainger et al.~\shortcite{G01}. We
subtract the sources from the original {\it uv}-data using point-source
models, implemented by the {\sc Aips} task {\sc Uvsub}.  The procedure
that we adopt for 
removing sources which are not point sources on our longest baselines, but are
not extended on the scale of the SZ decrement, is detailed
in Grainge et al~\shortcite{G96}.  If we find a bright source which is
extended on the scale of the SZ 
effect, we cannot be confident of subtracting the source accurately and
observations of the field have to be abandoned.

The SZ decrement is typically 
several arc-minutes in extent and so is resolved out
by the long-baseline antenna pairs, and to a good approximation is only
detected  by the short baselines.
We therefore either map only the short baseline source-subtracted data
(typically 0--1 k$\lambda$), or apply a {\it uv}-taper to
look for an SZ decrement close to the X-ray cluster centre.  The quality of
the resulting image is determined by both the structure of the SZ decrement
and the aperture-plane coverage of the observation. For high-declination
objects, e.g. Abell 2218~\cite{JonesCapri} or Abell 773~\cite{G93}, the
aperture-plane coverage is basically a circular ring at the radius of the
shortest baseline, with very little SZ signal at the second-shortest
baseline. An image made from this will show the position of the centre of the
decrement and reveal any significant departures from circular symmetry, but
does not strongly constrain the shape of the SZ profile. Additional
information, such as X-ray imaging and assumptions about the physics of the
gas, must be used to fully model the gas structure. In a low-declination
object such as Cl~0016+16 however, the projection of the baselines results in a
much better-filled aperture plane. Using two different array configurations
optimised for the north--south and east--west parts of the aperture plane
results in an image in which significant structure can be seen.  Since the SZ
signals generally have only moderate signal-to-noise ($\sim 5\sigma$),
{\sc Clean}ing to remove the point-spread-function 
has to be performed with great care.  With the poor aperture-plane
coverage used for the mapping, the inner part of the synthesised beam is
poorly represented by the Gaussian restoring beam, with the result that the
map's noise level will be artificially reduced, and that significant artifacts
can occur.  The best results are obtained by using a {\sc Clean \rm}-box
around the decrement, followed by a {\sc Clean} over the entire field, but
never cleaning deeper than about $2 \sigma$. 

{\sc Clean}ed maps of the SZ effect are useful for showing whether we
have detected 
any resolved structure in the decrement, and for overlaying on X-ray or
optical images to check for positional coincidence. However,
 it is necessary to do all analysis for
calculation of~$H_0$ in the aperture plane~\cite{KG01}, since the
noises are independent between visibilities, as opposed to the map plane
in which the pixel noises are strongly correlated.

\section{RT Observations of Cl~0016+16}

\begin{figure}
\begin{center}
\psfig{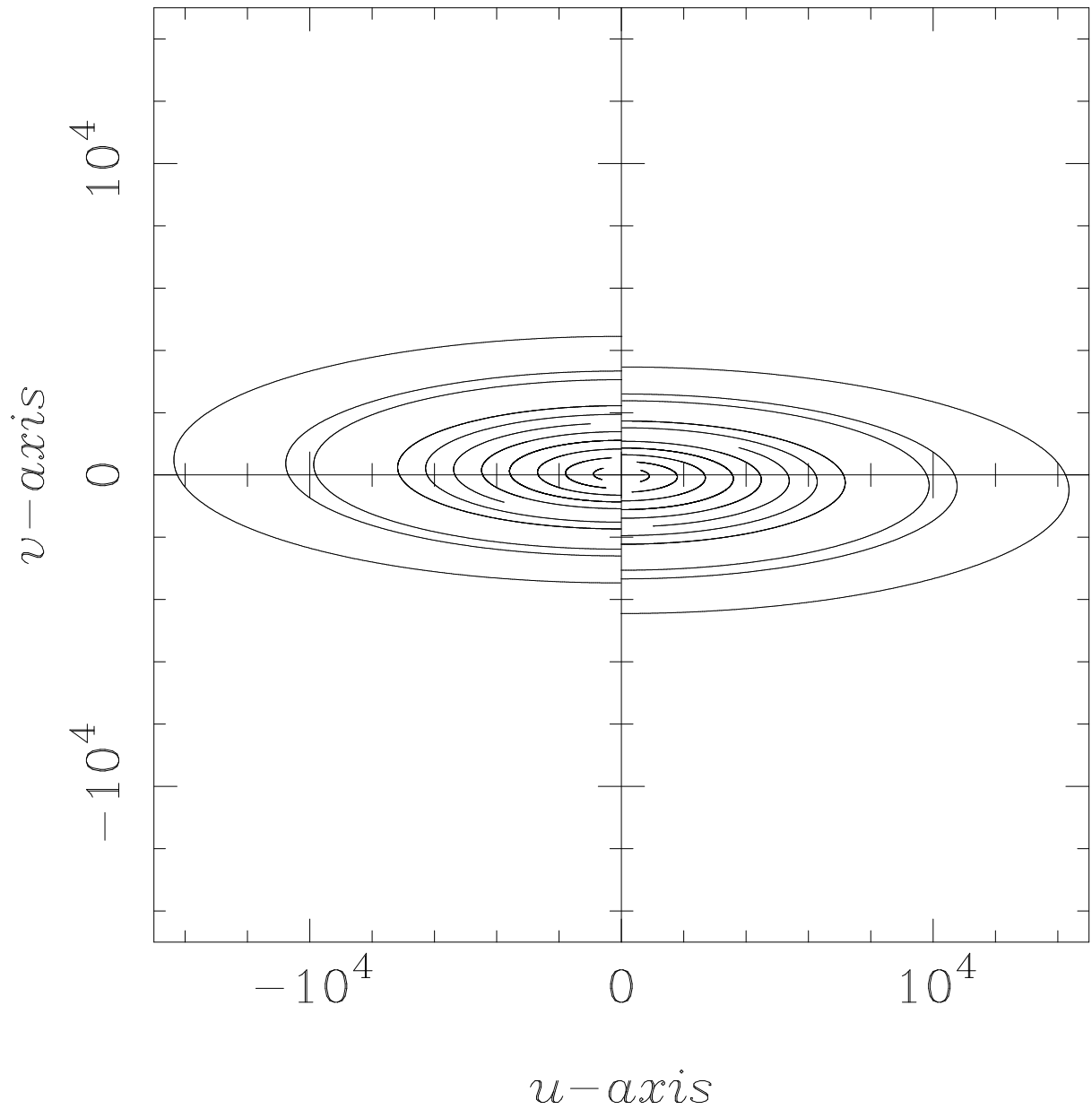}
\caption{Aperture-plane coverage for Cl~0016+16 observations. Two array
configurations were used, with relative antenna positions 0, 1, 2,
2.5, 3 and 0, 2.5, 4, 6, 8, where one unit is 35.6~m or 1820
wavelengths. Whenever one antenna is shadowed by another, all
baselines involving that antenna are discarded.\label{uv-coverage}}
\end{center}
\end{figure}

Two array configurations were used: the most compact array, Cb, which provides
short baselines at hour angles around zero, but in which the antennas shadow
each other at extreme hour angles, and a more extended array, Cd, to provide
unshadowed north--south baselines. Figure~\ref{uv-coverage} shows the
aperture-plane coverage. The asymmetry is due to the $\sim 3^{\circ}$ offset
of the baselines from east--west. The cluster was observed for the total of 46
days in the most compact array, and 32 days in the more extended array,
between 1993 July 20 and 1995 January 10 with simultaneous
observations of 0007+171 as a phase calibrator.
The pointing centre was $\rm 00^h 15^m 58^s.5 + 16^{\circ} 09^{\prime}
42^{\prime \prime}$ (B1950.0), which was also the phase centre.  Previous long
integrations with the RT have shown that it does not suffer from systematic
offsets at the phase centre.

\begin{figure}
\psfig{figure=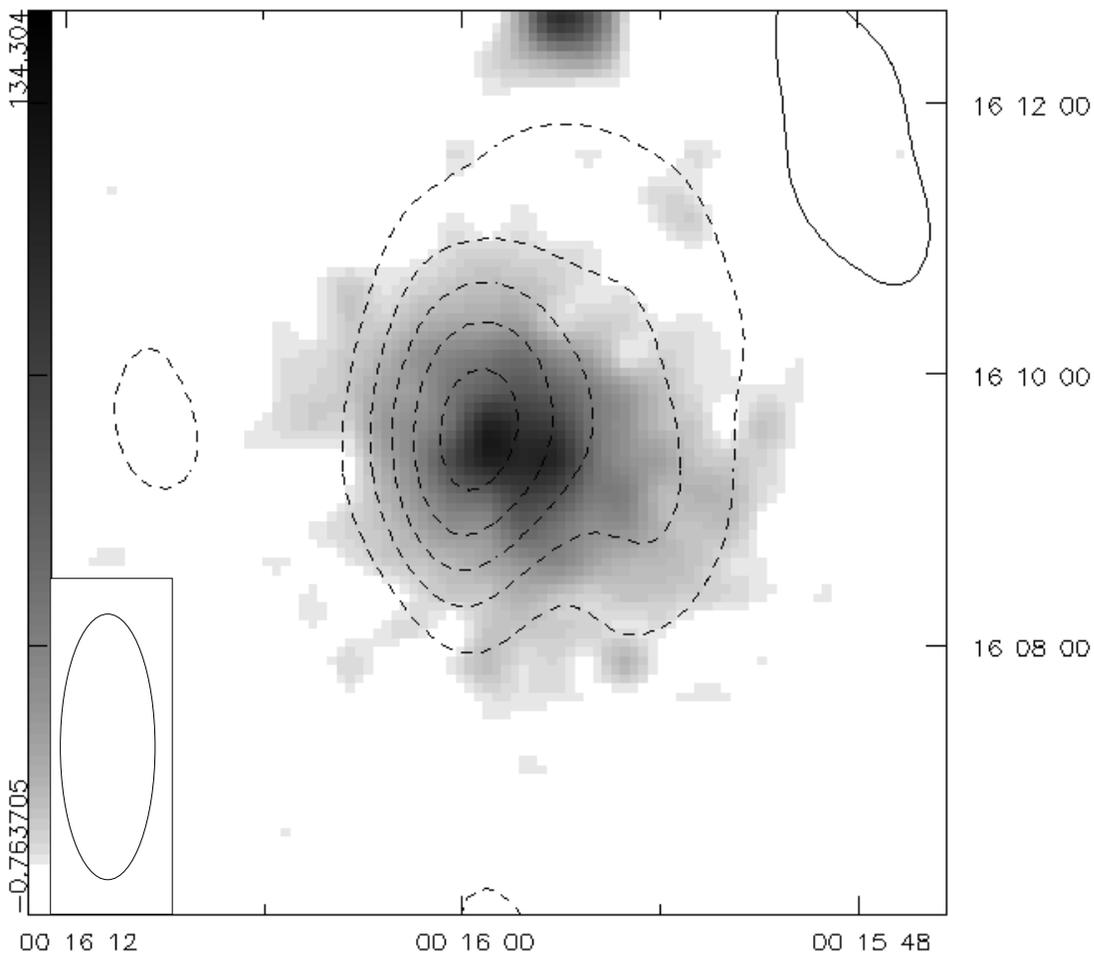, angle=270,width=15cm}
\caption{High resolution RT map of the SZ effect in Cl~0016+16
(contours) superposed 
onto the {\sc Rosat} PSPC image. The contours are -250, -200, -150,
-100, -50 and 50 $\mu$Jy. The restoring beam is
$143\times51^{\prime\prime}$ at an angle of $4.3^\circ$. The rms pixel
noise on the map is $40~\mu$Jy. The entire source-subtracted data set
has been used to produce this map, with a Gaussian taper with $1/e$
points at
2~k$\lambda$. 
\label{sz-hires}}
\end{figure}

\begin{figure}
\psfig{figure=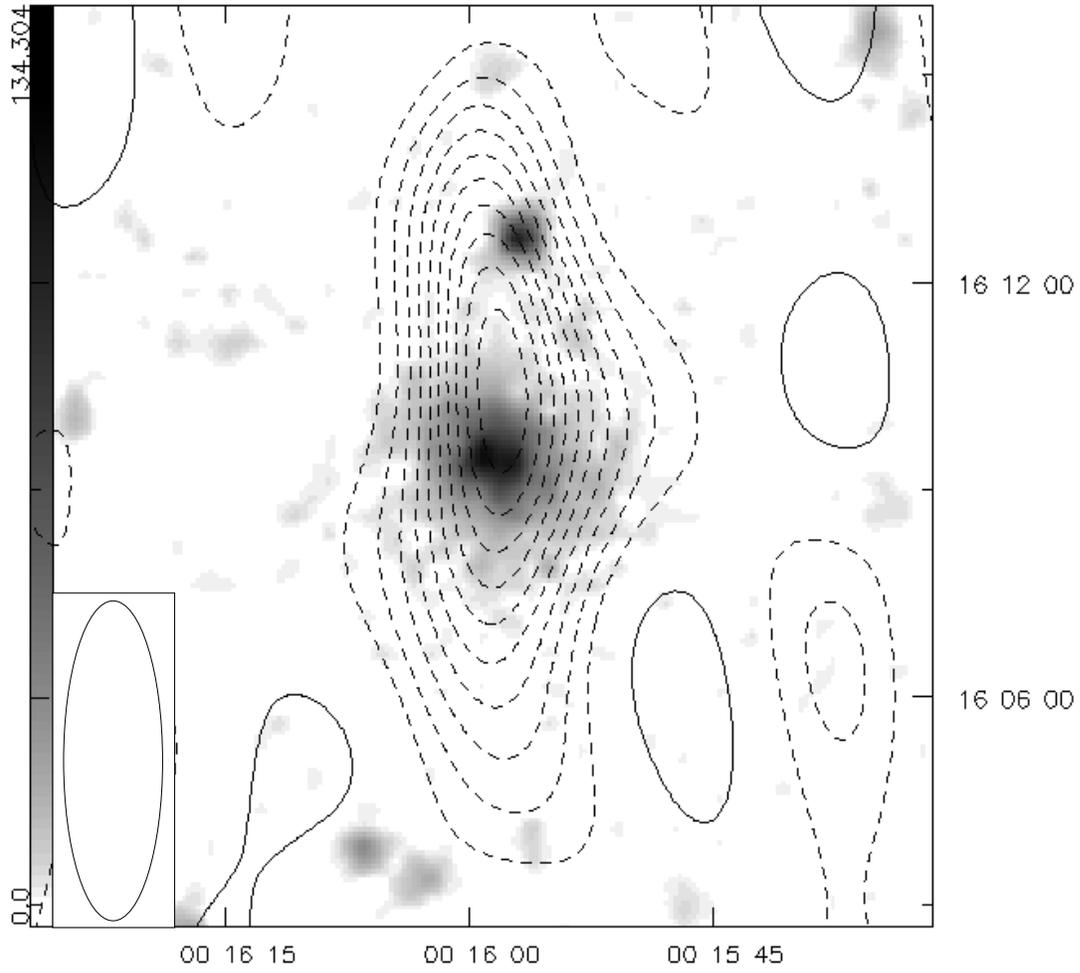, angle=270,width=15cm}
\caption{Low resolution RT map of the SZ effect in Cl~0016+16
(contours) superposed 
onto the {\sc Rosat} PSPC image. The contours are -800, -720, -640,
-560, -480, -400, -320, -240, -160, -80 and 80 $\mu$Jy. The restoring beam is
$333\times103^{\prime\prime}$ at an angle of $1.6^\circ$. The rms pixel
noise on the map is $90~\mu$Jy. 
Only data from the shortest RT baselines ($<870~\lambda$) have been
used in this map.
\label{sz-lores}}
\end{figure}

A map of the long-baseline data detected the presence of two sources
in the field. Both of
these correspond with sources in the survey of Moffet \&
Birkinshaw\shortcite{MB} (hereafter MB). The only other source within
$6^\prime$ (the RT primary beam) of our pointing centre that was
detected by MB is source MB14, the extended
cluster halo emission. This will be discussed further in
section~\ref{halo}. {\sc Fluxfitter} was run over the two array
configurations separately (because of the possibility of source
variability as discussed in Grainge et al.~\shortcite{G96}), 
and the results are shown in table~\ref{fluxes}.
Once these sources had been subtracted we mapped and {\sc Clean}ed the
entire data set at two different resolutions (Figures~\ref{sz-hires}
and~\ref{sz-lores}).
These maps show clear detections of the SZ decrement and the
high--resolution  image has a resolved 
east--west extension.
The position of the
peak and the extent of the SZ decrement agree well with the X-ray 
emission.

\section{Determining The Hubble Constant}

The method we use to measure $H_0$ from RT cluster data has been described in
detail by Grainge et al~\shortcite{KG01}. 
As in that paper we make our models  assuming an Einstein-de Sitter
cosmology, and then calculate corrections for other cosmologies.
Here we give details relevant to
Cl~0016+16. 

\subsection{X-ray image}\label{xfit}

We use the public {\sc Rosat} PSPC image of Cl~0016+16 (PI Hughes),
with live time of 43157~s,
 to fit for the shape of the gas density
distribution in the cluster
and to determine a value of $n_0^2/H_0$. 
We use the 0.5--2~keV band to minimise the effect of galactic emission and
absorption; for the later we assume a column density of $5.6\times
10^{20}\rm\, cm^{-2}$.
We fit to a two-dimensional
$\beta$-model  
 using Poisson statistics as the
number of counts per pixel is typically small, and excluding regions
containing X-ray point sources. We estimate the mean background level in a
$15^{\prime\prime}$ pixel to be $4.04\times 10^{-5}$ counts~$\rm s^{-1}$. 
We find best-fit parameters of $\beta=0.714$, core
radii of $40.8''$ and $32.9''$ at a position angle of $53^{\circ}$ 
and a central
electron density of $n_0=1.373\times 10^{4} \, h^{1/2} \rm ~m^{-3}$,
 where $H_0 = 100h \, \rm
km^{-1}Mpc^{-1}$,
assuming that the line-of-sight depth through the cluster is the
geometric mean of the two elliptical axes in the plane of the sky
and an X-ray emission constant of
$3.80\pm0.27 \times 10^{-69}$ $\rm counts~s^{-1}$ from 1~m$^3$ of gas
of electron density 1~m$^{-3}$ at a luminosity distance of 1~Mpc.
These are in
good agreement with the two-dimensional models
of Hughes \& Birkinshaw \shortcite{HB98}, who
use a very similar fitting method, and also with those of Neumann \&
B\"ohringer~\shortcite{NB97},  who fit with a $\chi^2$ statistic
after a Gaussian smoothing.

\begin{figure}
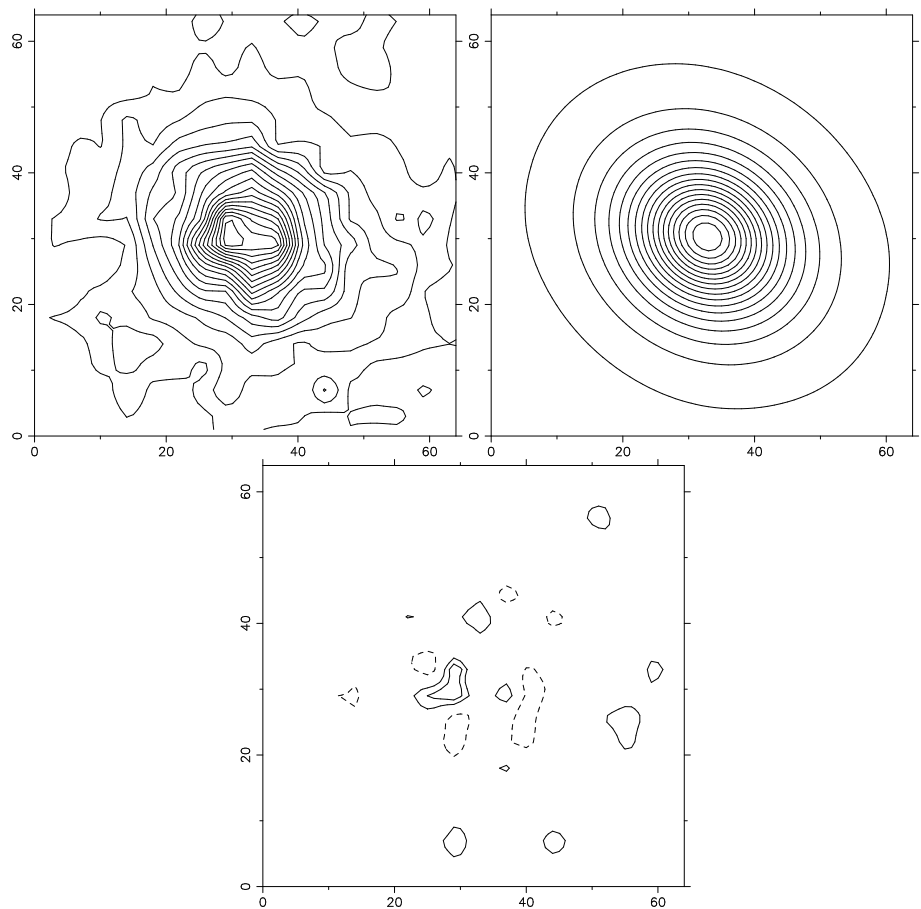

\begin{center}
\psfig{figure=rosat.eps, angle=270,width=6cm}
\psfig{figure=model_x.eps, angle=270,width=6cm}
\psfig{figure=resi.eps, angle=270,width=6cm}
\caption{The {\sc Rosat} PSPC image, our X-ray model,
and the residuals from the fit. Contours are every 0.4 counts per
$4^{\prime\prime}$ pixel.
\label{xfitmap}}
\end{center}
\end{figure}

Figure~\ref{xfitmap} shows the {\sc Rosat} PSPC image, our X-ray model,
and the residuals from our fit. 
The reduced value of $\chi^2$ for this fit is 0.945. However, since
the data are Poisson distributed with a relatively low
mean for many pixels, approximating the distribution to be Gaussian is
a poor assumption and therefore $\chi^2$ is not the correct statistic to
use.
We attempt to evaluate the goodness of fit by using the mean of the
log likelihoods of each pixel, given the model, over the image.
From our model we calculate 100 realisations of the image using the
Poisson distribution with the appropriate mean at each pixel, and find
the distribution of $L$ to be
${L}=
-2.339\pm0.034$. The corresponding value for the {\sc Rosat} image is
${L}= -2.32$. We therefore conclude that the fit is a good one.

\subsection{X-ray temperature}

We adopt the X-ray temperature calculated by Hughes \&
Birkenshaw~\shortcite{HB98}  of
$7.55^{+0.72}_{-0.58}$~keV. We assume the cluster to be isothermal at this
temperature. The effect of this probably inaccurate assumption on our
estimate of $H_0$ is discussed in Maggi et al.~\shortcite{M01}.

\subsection{Cluster halo emission} \label{halo}

Giovannini \& Feretti~\shortcite{GF00} 
have mapped the cluster halo source MB14 with the VLA in B, C and D
array at L-band~(1.4~GHz).  
These measurements give access to short enough baselines to ensure
that 
the full flux
density has been found and has not been resolved out.
They find that the total
integrated flux from this extended source is 5.5~mJy. Assuming that the
halo has a similar spectrum to that found in A665~\cite{JS96} we
estimate that the halo's flux density will have fallen to 55~$\mu$Jy by
15~GHz. 
We model the halo emission as a Gaussian with FWHM $2^\prime$ in order to 
estimate the extent to which we will resolve out the extended emission. We
find that only $51\pm9\%$ of the total zero-spacing flux is detected by the RT
shortest baseline of $870~\lambda$. 

The effect that halo emission will have on our calculation of $H_0$ is to
reduce the flux density of the detected SZ effect, which will lead to an
overestimate of $H_0$ by $(S_{halo} / S_{SZ})^2~\%$. 
Estimating that the SZ decrement detected by the RT has been reduced
by $30\pm30~\mu$Jy we apply a correction of $-7\pm7\%$ to our value of
$H_0$.

\subsection{$\bf H_0$ from Cl~0016+16}

\begin{figure}
\begin{center}
\psfig{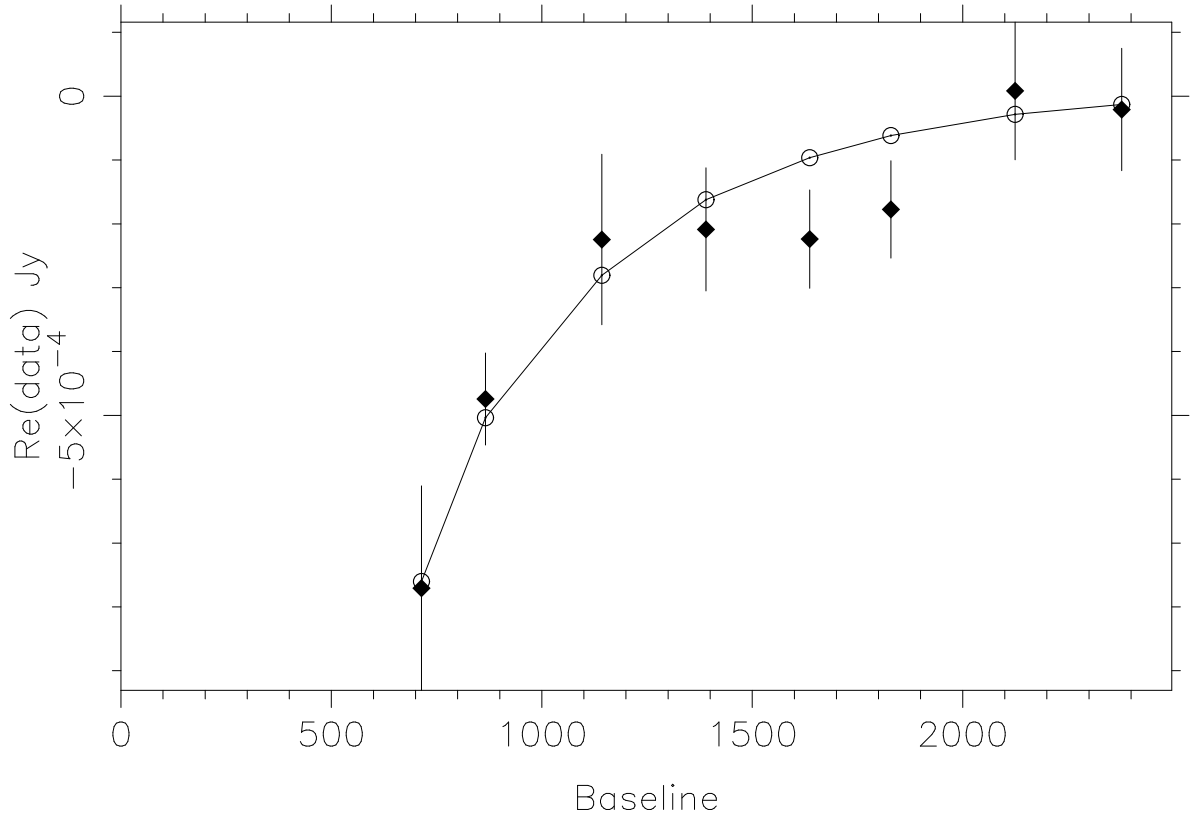}
\caption{
Binned real part of the source-subtracted visibilities (filled points
with error bars) 
for Cl~0016+16 observation 
against best fit model (open points joined by line). \label{fit}}
\end{center}
\end{figure}

\begin{figure}
\begin{center}
\psfig{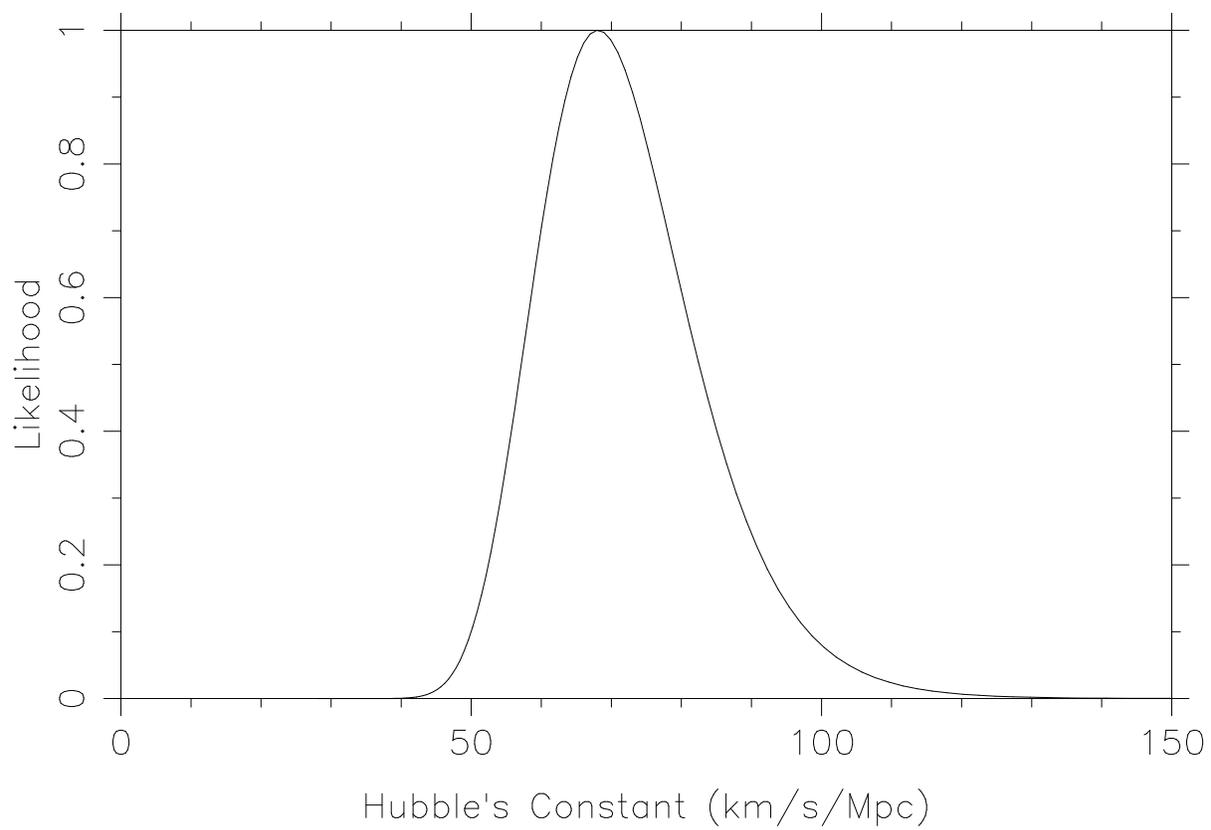}
\caption{Likelihood plot for values of $H_0$  from the fit of SZ decrement
to the X-ray derived cluster model. The width of the plot indicates
only the errors from the 
visibility data.
\label{likelihood}}
\end{center}
\end{figure}

Combining our X-ray model and SZ data in the aperture
plane we find a maximum likelihood fit of
$H_0=69^{+13}_{-9}~\rm km \ s^{-1} \ Mpc^{-1}$, $n_0=1.141\times 10^{3}\rm~
m^{-3}$, and a central decrement of $1077\pm 110~\mu \rm K$.
Figure~\ref{likelihood} shows the likelihood of the SZ data as a
function of $H_0$.
We find that the data fit very well the model predictions 
with a value of reduced $\chi^2=1.047$
and that the
SZ effect is detected with good significance on many
baselines~(Figure~\ref{fit}). 
Including the dominant sources of error (X-ray temperature, X-ray
fitting, SZ, ellipticity)~\cite{KG01} we find 
final estimates of  
$H_0=69^{+21}_{-16}~\rm km \ s^{-1} \ Mpc^{-1}$
 for a $\Omega_M=1.0$
cosmology; $H_0=84^{+25}_{-19}~\rm km \ s^{-1} \ Mpc^{-1}$
for $\Omega_M=0.3$ and $\Omega_\Lambda=0.7$;
and $H_0=76^{+23}_{-18}~\rm km \ s^{-1} \ Mpc^{-1}$
for $\Omega_M=0.2$ and $\Omega_\Lambda=0.0$

\section{Discussion}

Reese at el~\shortcite{R00} and Hughes and Birkinshaw~\shortcite{HB98} both
use Cl~0016+16 to calculate $H_0$. They find $63^{+12 \ +21}_{-9 \
-21}~\rm km \ s^{-1} \ Mpc^{-1}$ (in combination with data from MS 0451.6-0305;
assuming $\Omega_M=0.3, \ \Omega_\Lambda= 0.7$) and
$47^{+23}_{-15}~\rm km \ s^{-1} \ Mpc^{-1}$ (assuming $\Omega_M=0.2, \
\Omega_\Lambda= 0.0$) respectively. Many of the errors that go into
these estimates are identical to the ones that affect our
determination (e.g. the error on the cluster temperature). Therefore
the best way to check the consistency between these results is to
examine the various determinations of the SZ central decrement. These
are $1242\pm 105~\mu$K~(Reese et al.) and $1207\pm 190~\mu$K~(Hughes and
Birkinshaw), and so are somewhat higher than our value of
$1077\pm 110~\mu$K. 
Our result and that of Reese at el. differ by $1.1\sigma$
and so are not significantly discrepant, but we briefly 
discuss possible systematic differences between the two data
sets. Firstly it is possible that the halo
emission~(section~\ref{halo}) has a less steep spectrum than the value
we have adopted. In order to account fully for the difference we
calculate that the spectrum of the extended emission 
between 1.4 and 15~GHz
would have to be
approximately $\alpha^{15}_{1.4}=1.2$ ($S\propto\nu^{-\alpha}$). This
is flatter  than the flattest-spectrum halo emission seen in clusters
between 1.4 and 5~GHz of $\alpha^{5}_{1.4}=1.3$~\cite{H82}, and we
would expect the spectrum to steepen further between 5 and
15~GHz due to spectral aging (dominated by inverse Compton losses). 
Furthermore such a source would also affect the measurement of 
Hughes and Birkinshaw at 18~GHz to almost the same extent
as our observations at 15~GHz, but their
value is in very good agreement with Reese et al.
A second possibility is contamination from source M10. Reese et
al. do not remove this source from their  data but do estimate the
effect it could have by extrapolating from the 1.4~GHz flux and find
that its contribution is negligible.
We note that there is some evidence that this source is
variable with the result that estimating fluxes from observations at
different epochs is insecure, 
and that its spectral index between 1.4~GHz and 15~GHz 
is approximately 0.1, much flatter than the
assumed value of 0.7. 
We estimate that the effect of MB10 might be as much as $75~\mu$K at
30~GHz and that since this source lies in the negative sidelobes of
the point spread function with respect to the cluster
centre, its removal would reduce their calculated central decrement.

\section{Conclusions}

We have estimated the Hubble Constant by combining SZ and X-ray data
from the cluster Cl~0016+16 and find $H_0=69^{+21}_{-16}~\rm km \
s^{-1} \ Mpc^{-1}$ 
 for a $\Omega_M=1.0$
cosmology; $H_0=84^{+25}_{-19}~\rm km \ s^{-1} \ Mpc^{-1}$
for $\Omega_M=0.3$ and $\Omega_\Lambda=0.7$. This is in good agreement with
other recent determinations from a variety of methods:
the Hubble Key Project result of  
$H_0=72\pm8$~\cite{F00}; from a low-redshift sample of SZ clusters
$H_0=64^{+14}_{-11}$~\cite{MMR01}; from a high-redhift sample of SZ clusters 
$H_0=58^{+8}_{-7}$~\cite{J01}; 
from gravitational lensing  
$H_0=59^{+8}_{-7} \pm 15$~\cite{F99}  and
$H_0=69^{+13}_{-19}$~\cite{BB99};
and from combining cluster velocities, cosmic
microwave background and supernovae results
$H_0=74^{+14}_{-10}$~\cite{B01}.
Although optical data of Cl~0016+16 shows evidence
that it has 
undergone a recent merger event~\cite{Smail} we are able to fit a good
X-ray model to this cluster.
 This suggests that it should be possible to estimate the value
of the acceleration parameter, $q_0$, of the universe by combining SZ
and X-ray observations of a complete sample of high redshift ($z>0.5$)
clusters. This would be a useful check of the result from observations
of high redshift supernovae~\cite{P99}.

\subsection*{Acknowledgements}

We thank the staff of the Cavendish Astrophysics group who ensure the
continued operation of the Ryle Telescope. Operation of the RT is
funded by PPARC.

\clearpage

\begin{table*}
\begin{minipage}{150mm}
\begin{center}
\begin{tabular}{cccccl}
Configuration & Ae 1 & Ae 2 & Ae 3 & Ae 4 & Baselines Available\\ \hline
Ca & 12 & 9 & 7 & 4 & 2,3,3,4,5,5,7,8,9,12\\
Cb & 12 &10 & 8 & 4 & 2,2,4,4,4,6,8,8,10,12\\
Cc & 16 &12 & 8 & 4 & 4,4,4,4,8,8,8,12,12,16\\
Cd & 32 &24 & 16&10 & 6,8,8,10,14,16,16,22,24,32\\
\end{tabular}
\end{center}
\caption{\label{conf}Compact configurations of the Ryle Telescope.  The RT
antennas are numbered 1--8 from east to west, with antennas 1--4 being mounted
on a railway track. The convention used for specifying the parking positions
is to label them relative to the closest fixed antenna~5, in units of 9~m
($=450\lambda $).  There are stations every 36~m along the track at positions
4,~8,~12~... ~128, and additional ones at 7,~9 and~10. Antennas 6, 7 and 8 are
at positions $-128$, $-256$ and $-384$, but were not used in these
observations.}
\end{minipage}
\end{table*}

\clearpage

\clearpage

\begin{table}
\caption{Sources subtracted from Cl~0016+16 visibility data. Positions
are B1950. The errors in the fluxes are 39$\mu$Jy and 41$\mu$Jy in
arrays Cb and Cd respectively.}\label{fluxes}
\begin{tabular}{@{}ccccc}
RA & Dec & flux density & flux density & MB  \\
   &     &Cb array ($\mu$Jy)&Cd array ($\mu$Jy)&designation \\
00 15 56.2 & 16 04 04 & 417 & 187 & MB 15  \\
00 15 49.3 & 16 10 11 & 234 & 139 & MB 10  \\
\end{tabular}
\end{table}

\clearpage

\clearpage

\clearpage

\clearpage

\clearpage

\end{document}